\UseRawInputEncoding

\documentclass[10pt, prb, aps, twocolumn, showpacs, citeautoscript, floatfix, reprint, amsmath, amssymb, notitlepage, superscriptaddress]{revtex4-1}

\usepackage{graphicx}
\usepackage{stmaryrd}
\usepackage{rotating}
\usepackage{amsmath}
\usepackage{amsfonts}
\usepackage{amssymb}
\usepackage[countmax]{subfloat}
\usepackage{dcolumn} % align table columns on decimal point
\usepackage{bm} % bold math
\usepackage{color}

\usepackage{float}
\usepackage{latexsym}
\usepackage{wasysym}
\usepackage[pdfencoding=auto, psdextra]{hyperref}
\hypersetup{
    colorlinks,%
    citecolor=blue,%
    filecolor=blue,%
    linkcolor=blue,%
    urlcolor=blue
}
\usepackage[normalem]{ulem}

\begin{document}

\title{Proximity-induced zero-energy states indistinguishable from topological edge states}

\author{Igor J. Califrer}

\affiliation{Department of Physics, PUC-Rio, 22451-900 Rio de Janeiro, Brazil}

\author{Poliana H. Penteado}

\affiliation{Instituto de F\'{i}sica, Universidade Federal de Uberl\^{a}ndia, Uberl\^{a}ndia, Minas Gerais 38400-902, Brazil}

\author{J. Carlos Egues}

\affiliation{Instituto de F\'{i}sica de S\~ao Carlos, Universidade de S\~ao Paulo, 13560-970 S\~ao Carlos, S\~ao Paulo, Brazil}

\author{Wei Chen}

\affiliation{Department of Physics, PUC-Rio, 22451-900 Rio de Janeiro, Brazil}

\date{\today}

\begin{abstract}

When normal metals (NMs) are attached to topological insulators or topological superconductors, it is conceivable that the quantum states in these finite adjacent materials can intermix.  In this case -- and because the NM usually does not possess the same symmetry as the topological material -- it is pertinent to ask whether zero-energy edge states in the topological layer are affected by the presence of the NM. To address this issue, we consider three prototype systems simulated by tight-binding models, namely a Su-Schrieffer-Heeger/NM, a Kitaev/NM, and a Chern insulator/NM. 
For all junctions investigated, we find that there exist trivial ``fine-tuned'' zero-energy states in the NM layer that can percolate into the topological region, thus mimicking a topological state. These zero-energy states are created by fine-tuning the NM chemical potential such that some of the NM states cross zero energy; they can occur even when the topological material is in the topologically trivial phase, and exist over a large region of the topological phase diagram.
Interestingly, the true Majorana end modes of the Kitaev/NM model cannot be crossed by any NM state, as the NM metal layer in this case does not break particle-hole symmetry. 
On the other hand, when the chiral symmetry of the Su-Schrieffer-Heeger chain is broken by the attached NM, crossings are allowed. In addition, even in Chern insulators that do not preserve nonspatial symmetries, but the topological edge state self-generates a symmetry eigenvalue, such a fine-tuned zero-energy state can still occur. Our results indicate that when a topological material is attached to a metallic layer, one has to be cautious as to identify true topological edge states merely from their energy spectra and wave function profiles near the interface.

\end{abstract}

\maketitle

\section{Introduction}

The metallic edge or surface states are a defining feature of the topologically nontrivial phase of topological quantum materials, and often times what make them practically useful~\cite{Zhou08,Lu10,Hasan10,Qi11,Shen12,Bernevig13}. The pristine edge or surface states, hereafter collectively referred to as edge states, are responsible for a number of fascinating phenomena in these materials, such as the quantum spin Hall effect (QSHE)~\cite{Kane05,Kane05_2,Bernevig06,Bernevig06_2,Fu07} and Majorana fermions~\cite{Kitaev01,Oreg10,Lutchyn10}, among many others. 

A common strategy to access and probe edge states is to attach a thin metallic layer to a topological material such that the edge states percolate into the normal metal (NM). One then may be able to use these edge states to manipulate the metallic material. For instance, in 3D topological insulators (TIs), it has been demonstrated that a thin ferromagnetic metal (FMM) deposited on top of a TI exhibits very prominent current-induced spin torque~\cite{Mellnik14,Wang15_spin_torque_TI,Mahendra18}. This phenomenon has triggered significant theoretical interest in investigating the underlying mechanism and the role played by the edge states~\cite{Garate10,Yokoyama10,Mahfouzi12,Lu13,Sakai14,Fischer16,Mahfouzi16,Ho17,Ndiaye17,Okuma17_2,Ghosh18,
Laref20,Zegarra20_percolated_QSHE,Chen20_3DTIFMM}. 

In these kinds of junctions, specifically topological insulator/normal metal (TI/NM) or topological superconductor/normal metal (TSC/NM), with a finite size NM, the nature of the NM quantum-well states can elicit interesting issues. In particular, since the topological edge state can percolate into the NM, it is natural to speculate that the NM quantum-well states can similarly percolate into the TI layer. The question is then whether one can {\it unambiguously} distinguish these percolated NM quantum-well states from the true topological edge states by merely analyzing the profile of the wave function, which is often how one tries to image topological edge states by using some real space probe such as a scanning tunneling microscope (STM)~\cite{NadjPerge14,Feldman17,Kim18,PalacioMorales19,Liu20,Manna20}. 

As a concrete example, we mention perhaps the most controversial system involving (topological) zero-energy modes:  a 1D topological superconductor. In this system, a zero-energy mid gap mode gives rise to zero-bias conductance peaks (measured, e.g., via STM probes~\cite{NadjPerge14,Schneider21} or usual transport geometries~\cite{Mourik12, Das12,Deng12}) that, purportedly, signal Majorana zero modes. A number of recent theory papers~\cite{Vernek14,Cayao15,SanJose16,Fleckenstein18,Reeg18,Hess2021,Cayao21_2}  have pointed out that by fine tuning the system parameters, trivial zero-energy modes can emerge, thus making it challenging to discern true topological Majorana zero states from trivial ones. Earlier on, the authors of Ref.~\cite{Vernek14} had already shown that by fine-tuning the dot level and dot-chain coupling of a quantum dot side coupled to a Kitaev chain, the two-terminal conductance through the dot mimicked the $e^2/2h$ conductance peak of the topological case, even in the absence of {$p$}-type superconductivity (i.e., $\Delta = 0 $) [see their Fig. 3(d)]. 

More importantly, fine-tuning complicates matters further as zero modes can also arise from disorder effects, Andreev bound states, Kondo effect, etc., as is well known~\cite{Franz}. Recently, a protocol~\cite{protocol2,protocol} has been proposed to possibly identify these zero-energy modes through conductance measurements. Interestingly, Ref.~\cite{Hess2022} points out that trivial Andreev bands can emulate closing and reopening of bulk band gaps in the non-local conductance of nanowires, which could be detrimental to the protocol of Refs.~\cite{protocol2, protocol}.

In connection with the above zero-energy fine tuning issue, here we consider junctions of a topological material (a TI or a TSC) attached to or grown on a metallic material. We show that should one try to identify non-trivial edge states in these type of junctions by probing (i) the edge-state wave function profiles, e.g., whether it decays and only localizes in one sublattice (SSH model) and/or (ii) their corresponding zero-energy spectral features, e.g., via conductance measurements, then caution must be taken. This is so because in both TI/NM and TSC/NM junctions, it is possible to fine-tune the NM chemical potential such that a quantum well state has zero energy and a wave function profile that is indistinguishable from the true topological edge state. We investigate in detail the wave function profile of such a fine-tuned zero-energy quantum well state in several models of one- (1D) and two-dimensional (2D) TIs and TSCs, including the Su-Schrieffer-Heeger (SSH) model~\cite{Su79}, the Kitaev chain~\cite{Kitaev01}, and the Chern insulator, in both topologically trivial and nontrivial phases to show that it can exist over a large region of the topological phase diagram, with several different models for the adjacent NM. 

We organize the paper as follows. In Sec.~II A we start from the simplest 1D chiral-symmetric TI, namely the SSH model. We show that if chiral symmetry is locally preserved when a NM layer is attached to the SSH system, a zero-energy quantum well state emerges and its wave function profile is indistinguishable from a true topological edge state. We then investigate a Majorana chain/NM junction in Sec.~II B. Here, particle-hole symmetry ensures that the zero-energy states must appear in pairs. By fine-tuning the NM chemical potential $\mu_N$, we show that the wave functions of these pairwise states are, to some extent, indistinguishable from true Majorana fermions even when the system is topologically trivial. In Sec.~II C we study a 2D Chern insulator/NM junction. Our results indicate whether the zero-energy states at zero momentum display the same symmetry eigenvalue as the true topological chiral edge states depends on the orbital kinetics of the NM. Section III summarizes our results.

\section{Transient zero energy states in 1D and 2D}

In this section we investigate the SSH, the Kitaev, and the Chern-insulator models interfaced with a mesoscopic normal metallic layer. 

\subsection{Chiral symmetry: SSH/NM junction \label{sec:SSHNM}}

We first consider a 1D SSH/NM junction~\ref{fig:SSH_NM_junction}(a). Similar systems have been considered previously including an SSH/gapless wire/SSH junction~\cite{Dahan20}. Here, though, we will focus on an SSH/NM system with a finite NM layer\footnote{\label{footnote2} We emphasize that our metallic layer has a finite width, hence, a discrete set of quantum well-like states. Had we considered a semi-infinite metallic lead, i.e., a continuum of states, the self-energy due to the NM would simply broaden and shift the energy levels of the TI (TSC) layer.} such that the quantum well states in the NM couple only to one edge state. The spinless SSH layer of length $N_{TI}$ is described by the lattice Hamiltonian~\cite{Su79}
\begin{eqnarray}
H_{SSH}&=&\sum_{i<N_{TI},i\in{\rm odd}}\left(t+\delta t\right)\left(c_{i}^{\dag}c_{i+1}+c_{i+1}^{\dag}c_{i}\right)
\nonumber \\
&+&\sum_{i<N_{TI},i\in{\rm even}}\left(t-\delta t\right)\left(c_{i}^{\dag}c_{i+1}+c_{i+1}^{\dag}c_{i}\right),
\label{HSSH}
\end{eqnarray}
where $c_{i}$ ($c^\dagger_i)$ is the spinless fermion annihilation (creation) operator at site $i$, and $t\pm\delta t$ are the alternating hopping amplitudes. Figure~\ref{fig:SSH_NM_junction}(b) shows the spectrum of the SSH chain $H_{SSH}$ with open boundary conditions as a function of the parameter $\delta t$, from which the zero-energy edge states in the topologically nontrivial phase ($\delta t<0$) can be clearly identified.

For periodic boundary conditions and after performing a Fourier transform, the SSH Bloch Hamiltonian $H_{SSH}({\bf k})$ itself possesses time-reversal (TR), particle-hole (PH), and chiral symmetries. Denoting by $K$ the complex conjugate operator and $\hat {\sigma}=(\sigma_x,\sigma_y,\sigma_z)$ the Pauli-matrix vector, we can implement the symmetry operators for PH, TR and chiral symmetries by $T=K$, $C=\sigma_{z}K$, and $S=\sigma_{z}$, respectively, thus realizing a 1D class BDI model~\cite{Schnyder08,Kitaev09}. For the SSH model, chiral symmetry is often termed sublattice symmetry.

The NM layer of length $N_{NM}$ is modeled by the tight-binding Hamiltonian
\begin{eqnarray}
&&H_{NM}=\sum_{i>N_{TI}}\mu_{N}\,c_{i}^{\dag}c_{i}
\nonumber \\
&&+\sum_{N_{TI}<i<N_{TI}+N_{NM}}t_{N}\left(c_{i}^{\dag}c_{i+1}+c_{i+1}^{\dag}c_{i}\right),
\label{HNM}
\end{eqnarray}
with $t_{N}$ and $\mu_{N}$ the hopping parameter and chemical potential in the NM, respectively, as depicted in Fig.~\ref{fig:SSH_NM_junction} (a). Note that as long as $\mu_N$ is nonzero, the Bloch Hamiltonian $H_{NM}({\bf k})$ breaks PH and chiral symmetries. Therefore, $\mu_{N}$ serves as a parameter to control the symmetries of the NM. The two systems are coupled by a tunneling term given by $H_T = t'\left(c_{i}^\dag c_{i+1} + c_{i+1}^\dag c_{i} \right)$, where $t'$ is the hopping parameter between the last site on the SSH chain and the first site on the NM [see Fig.~\ref{fig:SSH_NM_junction} (a)].

For open boundary conditions, an SSH chain in the topologically nontrivial phase ($\delta t<0$) is known to host two zero-energy edge states that are eigenstates of the chiral symmetry operator $\sigma_{z}$~\cite{Shen12,Bernevig13}.
This translates into edge states localizing on odd and even sites near the left ($i\approx 1$) and right ($i\approx N_{TI}$) ends of the chain, respectively. 
When the NM layer is attached to the right end of the chain, the first question one can ask is how to unambiguously distinguish the percolated topological edge state from other eigenstates of the SSH/NM junction. To answer this question, we numerically diagonalize the SSH/NM junction to obtain its eigensolutions. We will discuss the results for the topologically nontrivial and trivial cases separately in the following sections.

\subsubsection{Results: SSH/NM junction in the nontrivial phase ($\delta t<0$)}

\begin{figure}[htb!]
\begin{center}
\includegraphics[clip=true,width=0.95\columnwidth]{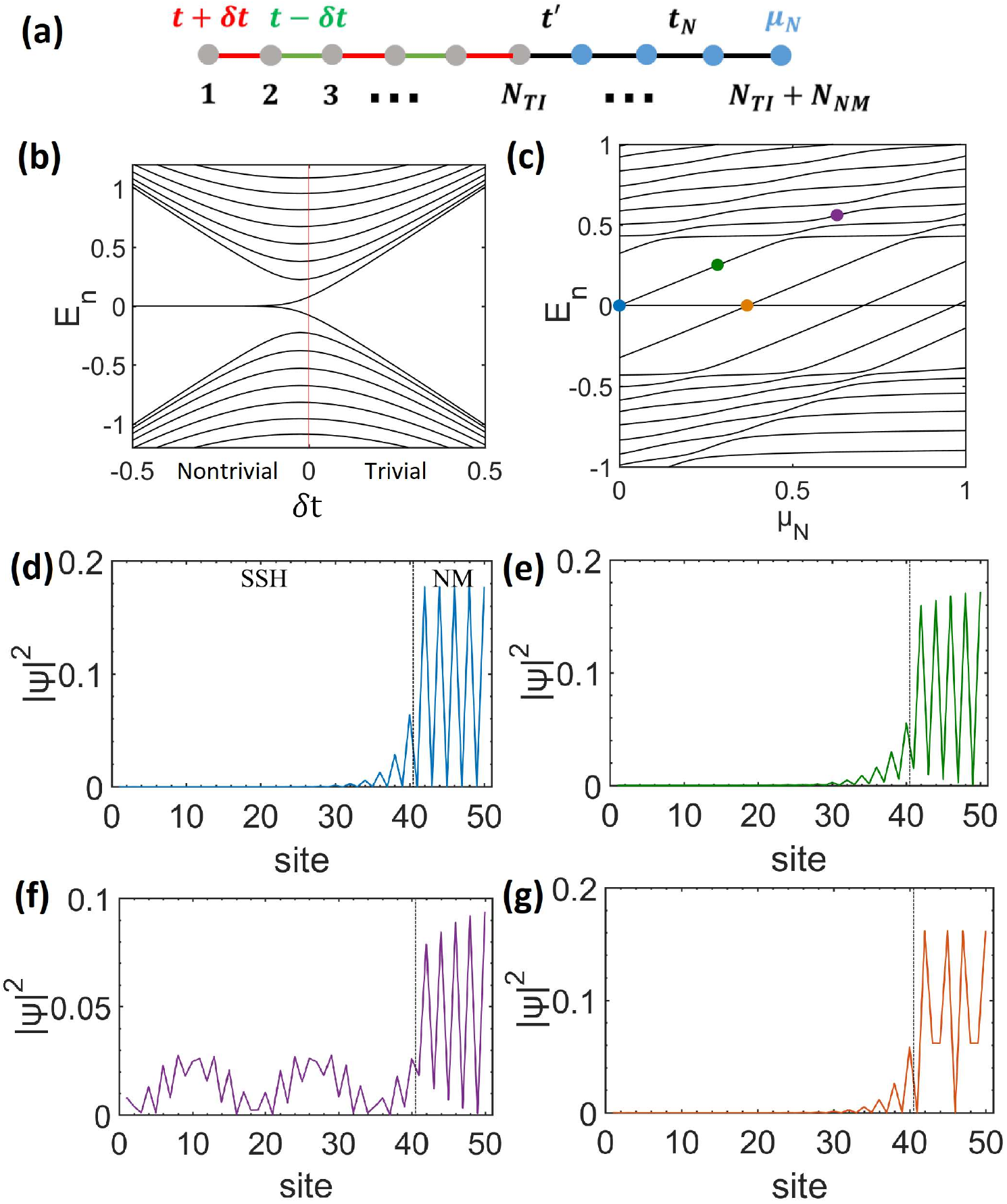}
\caption{(a) The 1D SSH/NM junction consists of an SSH chain of length $N_{TI}$ and kinetic hoppings $t\pm\delta t$, coupled to an NM layer of length $N_{NM}$, with chemical potential $\mu_N$ and hopping $t_{N}$, via an interface hopping $t'$. (b) Eigenenergies $E_n$ (measured in units of $t$) for an isolated SSH chain as functions of $\delta t$. (c) Evolution of the eigenenergies $E_n$ in the topologically nontrivial phase ($\delta t <0)$ for the SSH/NM junction in (a) as the chiral-symmetry-breaking chemical potential $\mu_{N}$ is increased. (d) to (g) wave function profiles of several eigenenergies in (c) (following the same color code): (d) $\mu_N=0$ (blue solid circle), (e) $\mu_N= 0.2844t$ (green solid circle), (f) $\mu_N=0.6281t$ (purple solid circle), and  (g) $\mu_N=0.3656t$ (orange solid circle). We have used the following parameters: $N_{TI}=40$, $N_{NM}=10$, ${\delta t}=-0.2 t$, $t'= t$, and $t_N = -0.6t$. }
\label{fig:SSH_NM_junction}
\end{center}
\end{figure}

Figure~\ref{fig:SSH_NM_junction} summarizes our results for the topological regime of the SSH chain.  In Fig.~\ref{fig:SSH_NM_junction}(c) we show the discrete eigenenergies $E_n$ of the system as a function of the chiral-symmetry-breaking NM chemical potential $\mu_{N}$. When $\mu_{N}=0$, the state at zero energy can be identified as a true edge state [see solid blue circle in Fig.~\ref{fig:SSH_NM_junction}(c)] because its wave function percolates into the NM layer and retains the SSH sublattice symmetry, i.e., it only localizes on the even sites, see Fig.~\ref{fig:SSH_NM_junction}(d). As we gradually turn on $\mu_{N}$, the energy of this edge state increases, thus allowing us to track its evolution. The wave function at finite $\mu_{N}$ no longer entirely localizes on the even sites; it continuously evolves into a sublattice symmetry-breaking profile, see Fig.~\ref{fig:SSH_NM_junction}(e), corresponding to the green solid circle in Fig.~\ref{fig:SSH_NM_junction} (c). At an even larger $\mu_{N}$, such that the tracked edge-state energy merges into the bulk (see purple solid circle), its identification becomes rather ambiguous, see Fig.~\ref{fig:SSH_NM_junction}(f).

Figure \ref{fig:SSH_NM_junction}(c) also reveals that tailor-made zero-energy states appear as we vary $\mu_{N}$. This occurs because some negative energy NM states at $\mu_N=0$ gradually move up as we increase $\mu_N$, eventually crossing zero energy, e.g., the orange solid circle. The wave function of this kind of state inherits both the edge state character (localization on even sites within the SSH layer), and features of the NM quantum well-like states (standing waves within the NM layer), as shown in Fig.~\ref{fig:SSH_NM_junction}(g). Thus, if a local measurement, like STM, is performed in the SSH/NM interface, it would not be possible to distinguish these {\it fine-tuned} zero-energy states~\ref{fig:SSH_NM_junction}(g) from the true topological edge states~\ref{fig:SSH_NM_junction}(d).

\subsubsection{Results: SSH/NM junction in the trivial phase ($\delta t>0$)}

\begin{figure}[htb!]
\begin{center}
\includegraphics[clip=true,width=0.95\columnwidth]{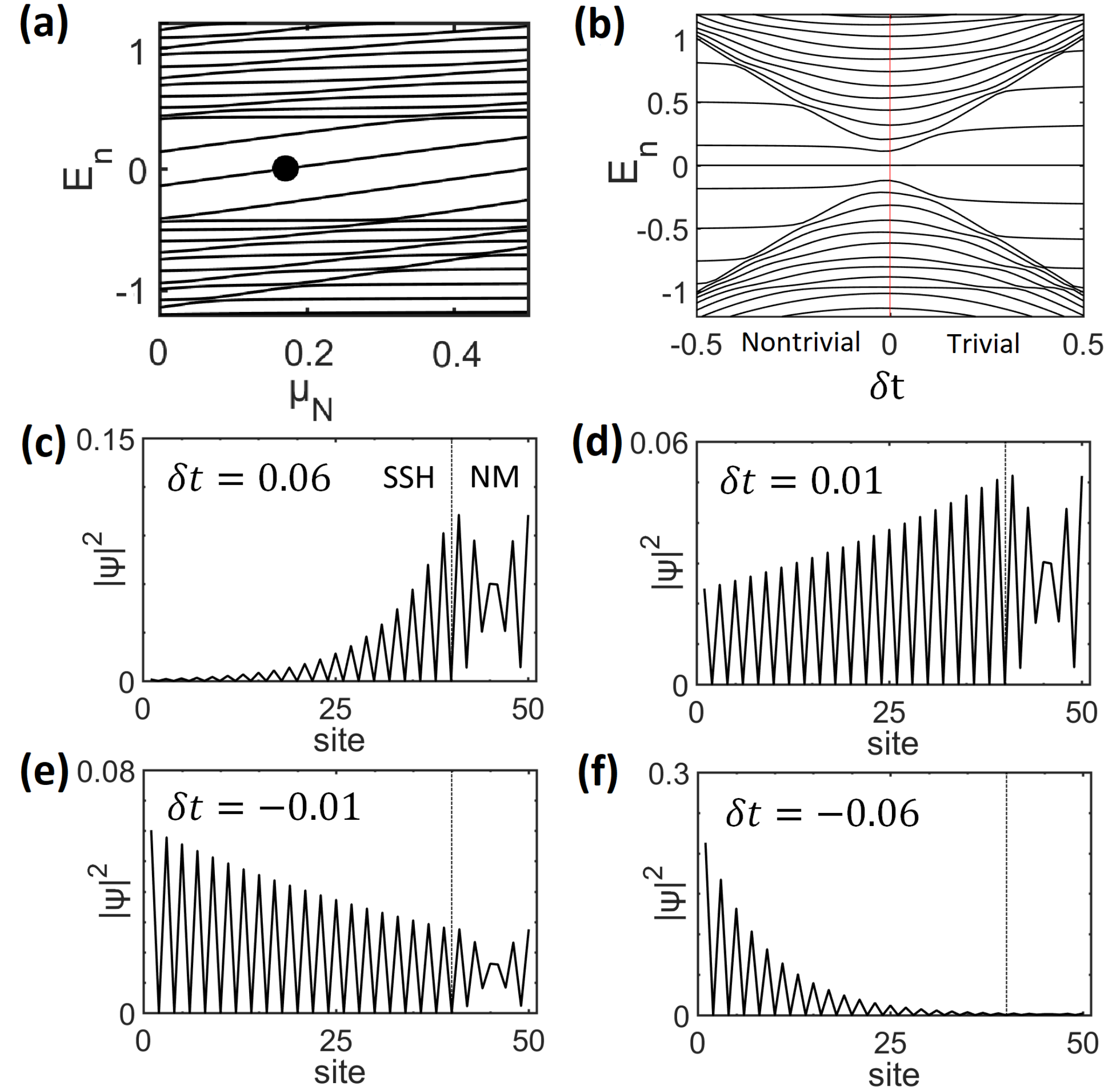}
\caption{(a) Eigenenergies $E_n$ as functions of $\mu_{N}$ in the topologically trivial phase of the SSH chain ($\delta t =0.2 t$). (b) Eigenenergies for a fixed value of $\mu_N=0.1704t$ [black point in (a)] as functions of the hopping $\delta t$. Note that if cross the predicted trivial/nontrivial bulk phase transition ($\delta t =0$), there is no distinction between the two regimes. Wave function probability density for several $\delta t$ of the fine-tuned zero-energy state that only localizes on the even sites within the SSH layer and whose profile is indistinguishable from the true topological edge state: (c) $\delta t = 0.06t$, (d) $\delta t = 0.01t$, (e) $\delta t = -0.01t$, and (f) $\delta t = -0.06t$   We have used the following parameters: $N_{TI}=40$, $N_{NM}=10$, $t'= t$, $t_N = -0.6t$, and $\mu=0.4t$.}
\label{fig:SSH_NM_junction-trivial}
\end{center}
\end{figure}

When the SSH model is in the topologically trivial phase, which has no topological zero-energy edge states, one can still fine-tune the chemical in the metallic layer such that the NM quantum well states have zero energy, see Fig~\ref{fig:SSH_NM_junction-trivial}(a). The robustness of this fine-tuned mode can be seen in Fig.~\ref{fig:SSH_NM_junction-trivial}(b), in which we fix the NM chemical potential $\mu_N$ and vary the parameter $\delta t$ that controls the topology. As we move across the trivial/non-trivial bulk phase transition ($\delta t = 0$), the fine-tuned zero-energy state still persists, indicating that this state can exist over a wide region of the topological phase diagram parametrized by $\delta t$. The wave function probability density for different $\delta t$'s is shown in Figs.~\ref{fig:SSH_NM_junction-trivial}(c)--(f). We note that these states still localize only on the even sites within the SSH layer, meaning that they are indistinguishable from the true topological edge states.

Our result suggests that if one solely relies on the existence of the edge state to judge whether the system is in a topologically nontrivial phase, then caution must be taken in these kinds of SSH/NM junctions. 

\subsection{PH symmetry: Kitaev/NM junction}
\label{Kitaev/NM}

We now consider the spinless Kitaev $p$-wave superconducting chain interfaced with an NM layer. Similarly to the SSH model, the finite Kitaev model hosts topologically zero-energy end modes known as Majorana bound states~\cite{Kitaev01}.Note that a similar interface problem of a Majorana fermion leaking into an adjacent quantum dot has been investigated previously from the transport point of view~\cite{Vernek14,RuizTijerina15,Hoffman17,Schuray17}, as well as the situation that the two ends of the NM (that may also contain Rashba spin-orbit coupling) are coupled to two separated Majorana chains~\cite{Dahan17,Cayao21}. In contrast, here we focus on the symmetry perspective of a Majorana end mode leaking into a metal of finite length.  

The lattice Hamiltonian of this Kitaev/NM junction is given by
\begin{eqnarray}
H&=&\sum_{i\in N_{p}}t\left(c_{i}^{\dag}c_{i+1}+c_{i+1}^{\dag}c_{i}\right)-\mu \sum_{i\in N_{p}}c_{i}^{\dag}c_{i}
\nonumber \\
&+&\sum_{\left\{i,i+1\right\}\in N_{p}}\Delta \left(c_{i}c_{i+1}+c_{i+1}^{\dag}c_{i}^{\dag}\right)
\nonumber \\
&+&\sum_{i \in N_{NM}}t_{N}\left(c_{i}^{\dag}c_{i+1}+c_{i+1}^{\dag}c_{i}\right)-\mu_{N}\sum_{i\in N_{NM}}c_{i}^{\dag}c_{i} \nonumber \\
&+& t'\left( c_{N_p}^\dag c_{N_p+1} + c_{N_p+1}^\dag c_{N_p}\right),
\label{MajoranaNM_junction_lattice_model}
\end{eqnarray}
where $c_{i}$ ($c_i^\dag$) is the spinless fermion annihilation (creation) operator at site $i$ and $N_{p}$ ($N_{NM}$) denotes the Majorana chain (NM region). The parameters $t$, $\mu$, and $\Delta$ correspond to the hopping, chemical potential, and superconducting pairing in the Kitaev model, respectively. The last term in Eq.~\eqref{MajoranaNM_junction_lattice_model} corresponds to the coupling between the Kitaev and NM chains. The site indices $N_p$ and $N_{p}+1$ denote, respectively, the end of the Kitaev chain and the beginning of the NM layer, i.e., their interface.

\begin{figure}[htb!]
\begin{center}
\includegraphics[clip=true,width=0.99\columnwidth]{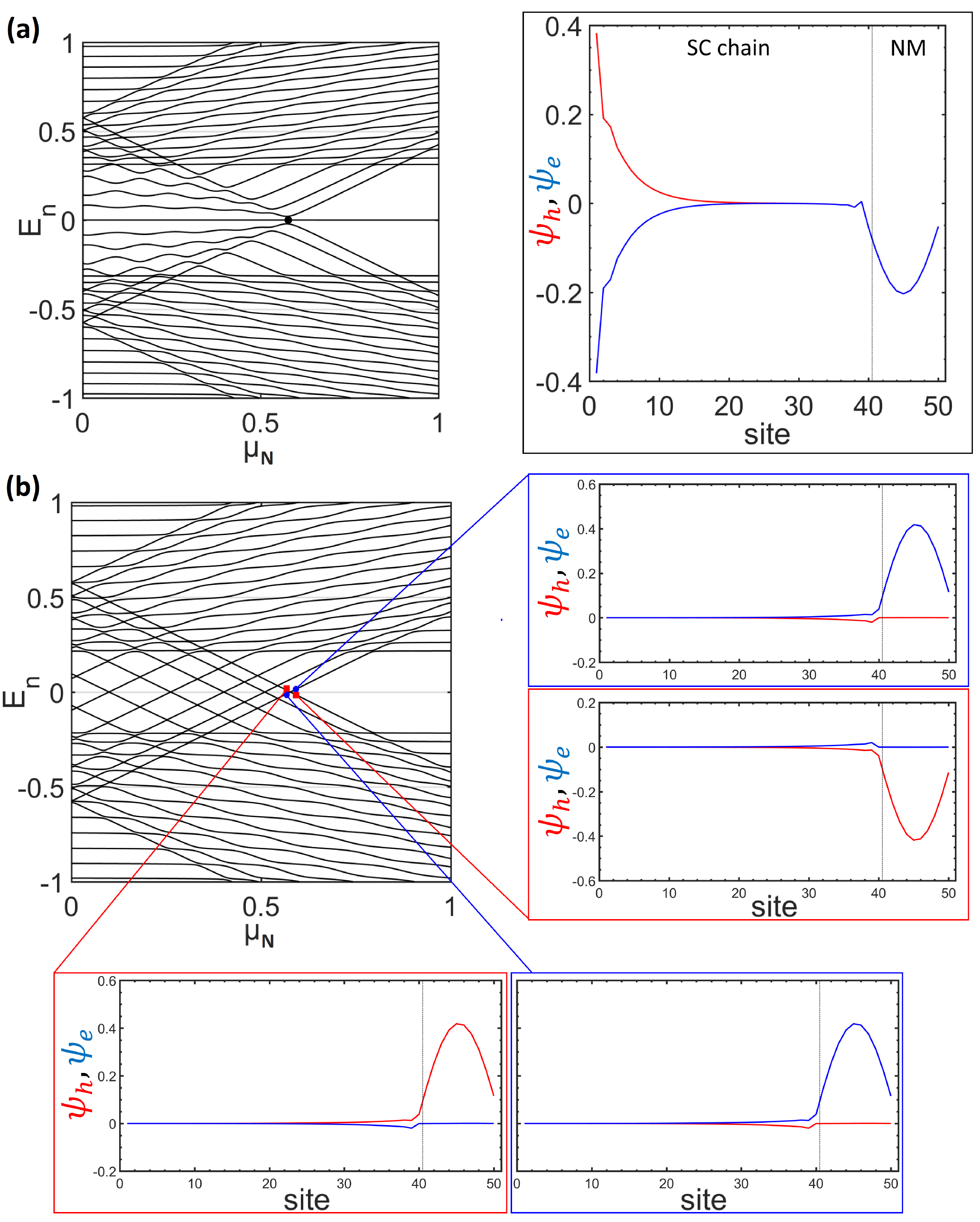}
\caption{ (a) Energy spectrum $E_n$ of a Kitaev/NM junction in the topologically nontrivial phase as a function of the NM chemical potential $\mu_{N}$. Here $N_{TI}=40$, $N_{NM}=10$, $\mu=0.5$, $t=0.4$, $\Delta=0.6$, and $t'= t_N=0.3$. Inset: wave function profiles of the two zero-energy states, indicating that one of them has a Majorana wave function $\psi_{e}=-\psi_{h}$ (red opposite of blue) at the left end, and the another has a Majorana wave function $\psi_{e}=\psi_{h}$ (red equals to blue, hence only blue is shown) at the right end that percolates into the NM. (b) Energy spectrum of the junction in the topologically trivial phase, $\mu=1$. The four panels show the wave function of the two quantum well states that cross zero energy upon tuning $\mu_{N}$. One state has only an electron-like component while the other has only a hole-like component, and hence, the two of them together may be misidentified as one Majorana fermion.}
\label{fig:Kitaev_NM_junction}
\end{center}
\end{figure}

\subsubsection{Results: Kitaev/NM junction in the nontrivial phase ($|\mu|<2t$) \label{Kitaev:nontrivial}}

The Bloch Hamiltonian of the Majorana chain, written in the basis $(c(k),c^{\dag}(-k))^{T}$, is given by
\begin{eqnarray}
H_{p}(k)=\left(\begin{array}{cc}
\frac{1}{2}\left(2t\cos k-\mu_{p}\right) & i\Delta\sin k \\
-i\Delta\sin k & -\frac{1}{2}\left(2t\cos k-\mu_{p}\right)
\end{array}\right),\;\;\;
\label{Kitaev_Hp_k}
\end{eqnarray}
which satisfies PH symmetry implemented by $C=\sigma_{x}K$. The NM chain alone expressed in the same basis is essentially the diagonal part of Eq.~(\ref{Kitaev_Hp_k}), which also satisfies PH symmetry. As a result, the edge state of the Kitaev/NM junction still remains an eigenstate of the PH-symmetry operator $C$ and therefore contains equal weights of the electron and hole channels. This means that the edge state can percolate into the NM layer and remain a Majorana fermion, as can be seen in Fig.~\ref{fig:Kitaev_NM_junction} (a), where the red (hole) and blue (particle) merge together to become one blue line that percolates into the NM.

One may consider this result as an artifact stemming from the Hamiltonian being expressed in the basis $(c(k),c^{\dag}(-k))^{T}$, so the PH symmetry by construction cannot be broken even when the system is attached to an NM layer. In addition, the energy spectrum shown in Fig.~\ref{fig:Kitaev_NM_junction} (a) indicates that even by fine-tuning the NM chemical potential $\mu_{N}$, NM-layer states never reach zero energy, so it is not possible to fabricate a zero-energy state in the topologically nontrivial phase other than the true Majorana edge state itself. This follows from our spinless system being one-dimensional and having PH symmetry. Therefore, it allows for at most two solutions per energy. \footnote{\label{footnote1} One observes that in the SSH/NM junction, chiral symmetry is broken by the presence of the NM layer when $\mu_N \neq 0$, and this allows the zero-modes to move away from zero energy. As it turns out, the state localized on the left edge (odd sites), far away from the interface, remains at zero energy since it is only weakly coupled to the NM. The state close to the interface, on the other hand, is shifted away from zero, which opens up the possibility of additional crossings at zero energy (at most two per energy), as shown in Fig.~\ref{fig:SSH_NM_junction} (b). In contrast, in the Kitaev/NM junction PH symmetry is not broken by the NM layer. Hence, in the topological regime, no states are allowed to cross at zero energy. Here the only two possible solutions are the true topological Majorana zero modes. } 

%{\cblue (1) Wei: Carlos suggests that: "we note that if the two Majorana modes interact, say with energy $\varepsilon_m$, they split and become Andreev bound states at energies $\pm \varepsilon_m/2$. In this case, crossings are allowed." Let's discuss if we should check this. }

\subsubsection{Results: Kitaev/NM junction in the trivial phase ($\mu<-2t$ or $\mu>2t$) \label{Kitaev:trivial}}

In the topologically trivial phase (here we only consider $\mu>2t$), in which no Majorana edge states occur, the situation is different. In contrast to the topological case in Sec.~\ref{Kitaev:nontrivial}, here, the NM quantum well states can be made to cross at zero energy by fine-tuning $\mu_N$, as shown in Fig.~\ref{fig:Kitaev_NM_junction}(b).
In our system, this is possible because there are no topological zero-energy solutions. Since the spectrum is PH symmetric, the states always come in pairs with one electron-like and the other hole-like.
In the wave-function panels of Fig.~\ref{fig:Kitaev_NM_junction} (b), we show the wave-function profile before ($\mu_{N}=0.5672$) and after ($\mu_{N}=0.5922$) the crossing. Note that the inversion of the electron-like to hole-like at the crossing is clearly visible. 

\begin{figure}[htb!]
\begin{center}
\includegraphics[clip=true,width=0.99\columnwidth]{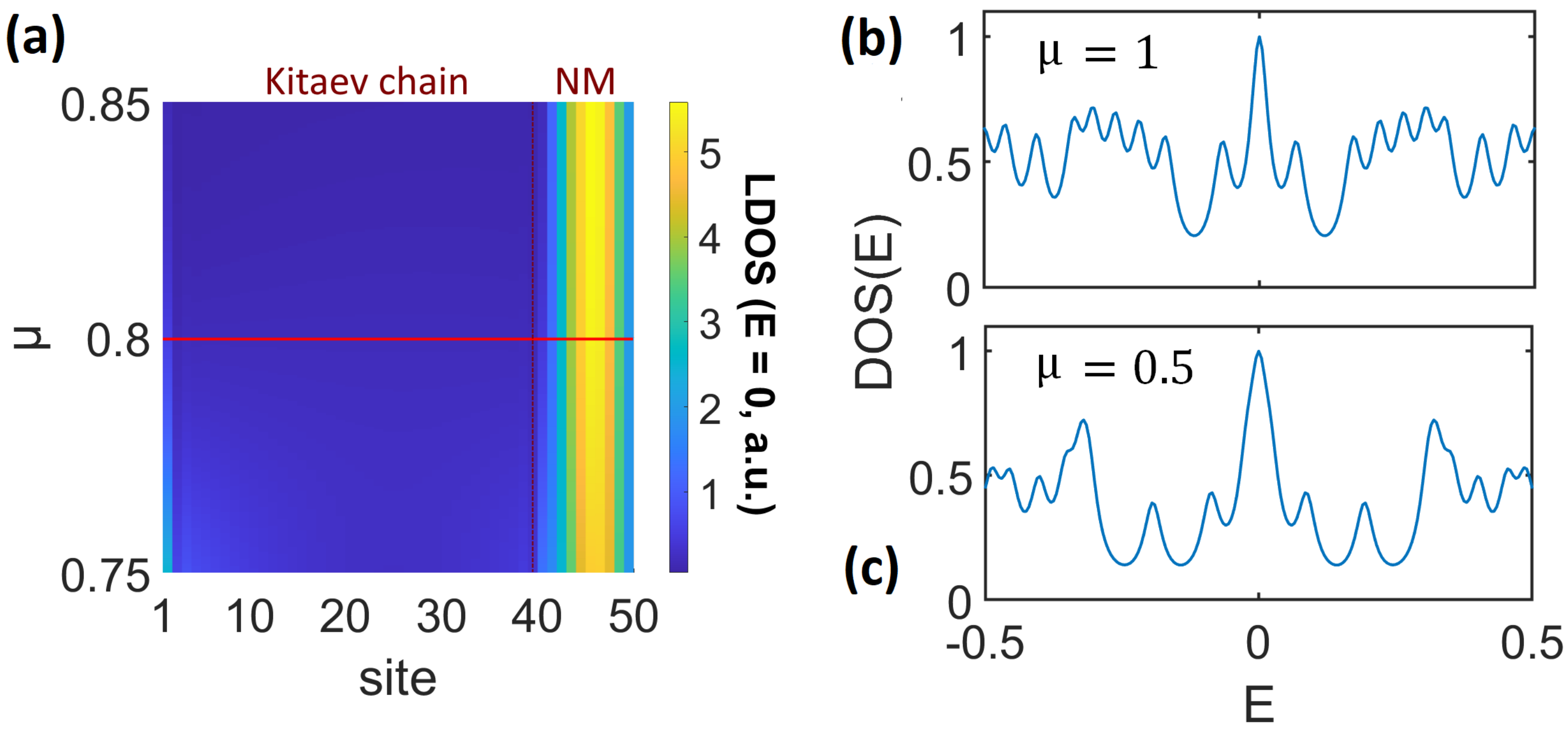}
\caption{(a) Color map of the LDOS at $E=0$ for the Kitaev/NM junction as a function of $\mu$ and the number of sites. The horizontal red line indicates the topological ($\mu<2t$)/non-topological ($\mu>2t$) phase transition. Total density of states for the (b) trivial ($\mu=1$) and (c) topological ($\mu=0.5)$ regimes as functions of the energy $E$.}
\label{fig:Kitaev_NM_DOS}
\end{center}
\end{figure}

As pointed out in a different context by recent theoretical and experimental works~\cite{Liu2017, Moore2018,Avila2019,Woods2019,Vuik2019,Chen2019,Pan2020,Valentini2021,Hess2021}, this feature raises concerns that either of these two zero-energy states (trivial and nontrivial) may be mistakenly identified as a Majorana mode. Additionally, we corroborate this difficulty by showing in Fig.~\ref{fig:Kitaev_NM_DOS} (a) the local density of states (LDOS) at $E=0$ of the Kitaev/NM junction as a function of the Kitaev chemical potential $\mu$ and the number of sites. The horizontal red line indicates the predicted topological/non-topological bulk phase transition. We note that throughout the whole NM region, there is an energy mode pinned to zero across the topological phase transition, which can lead to an ambiguity in telling apart the true Majorana mode. 

To further support this statement, Figs.~\ref{fig:Kitaev_NM_DOS} (b) and (c) show the density of states (integral of the LDOS) for the trivial and nontrivial regimes, respectively, as functions of the energy $E$. Once again, we see that they both have a zero bias peak and very similar profiles, indicating that tunneling conductance measurements would possibly not distinguish them.

\subsection{Symmetry induced by edge direction: Chern insulator/NM junction}

The Chern insulator belongs to class A in the Altland$-$Zirnbauer symmetry classification~\cite{Schnyder08}, which has no nonspatial symmetries. Despite the absence of nonspatial symmetries, the edge state self-generates a symmetry eigenvalue according to the direction of the edge, as we shall see below. 

The bulk Chern insulator is described by the Bloch Hamiltonian\cite{Qi11,Bernevig13} 
\begin{eqnarray}
H_{CI}({\bf k})&=&A\sin k_{x}\sigma^{x}+A\sin k_{y}\sigma^{y}
\nonumber \\
&+&\left(M+4B-2B\cos k_{x}-2B\cos k_{y}\right)\sigma^{z},
\label{HCh_kspace}
\end{eqnarray} 
written in the spinless basis $\psi=\left(c_{{\bf k},s},\;c_{{\bf k},p}\right)^{T}$, where $c^\dagger_{{\bf k},s(p)} = \sum_ie^{-i{\bf k}\cdot{\bf r}} c^\dagger_{i,s(p)}$, and $c^\dagger_{i,s(p)}$ is the creation operator of an $s(p)$-orbital electron at site $i$. Recall that in~\eqref{HCh_kspace} the Pauli matrix vector $\hat{\sigma}=(\sigma_x,\sigma_y,\sigma_z)$ denotes a pseudospin degree of freedom. The corresponding 2D lattice model~\cite{Chen20_absence_edge_current} reads 
\begin{eqnarray}
&&H=\sum_{i\in CI}t\left\{-ic_{i,s}^{\dag}c_{i+a,p}
+ic_{i+a,s}^{\dag}c_{i,p}+h.c.\right\}
\nonumber \\
&&+\sum_{i\in CI}\left\{-c_{i,s}^{\dag}c_{i+b,p}+c_{i+b,s}^{\dag}c_{i,p}+h.c.\right\}
\nonumber \\
&&+\sum_{i\in CI,\delta}t'\left\{-c_{i,s}^{\dag}c_{i+\delta s}+c_{i,p}^{\dag}c_{i+\delta, p}+h.c.\right\}
\nonumber \\
&&+\sum_{i\in CI}\left(M+4t'\right)\left\{c_{i,s}^{\dag}c_{i,s}
-c_{i,p}^{\dag}c_{i,p}\right\},\;\;\;\;
\label{Hamiltonian_2DclassA}
\end{eqnarray} 
with $CI$ denoting the Chern insulator sites, $t=A/2$, $t'=B$, $s$ and $p$ the orbital degrees of freedom, and $\delta=a, b$ the lattice constants for the ${\hat{\bf x}}$ and ${\hat{\bf y}}$ directions, respectively. Periodic boundary conditions along ${\hat{\bf x}}$ and open boundary conditions in the ${\hat{\bf y}}$ direction are imposed in our calculations.

We focus on the critical region near $M=0$ where the bulk gap closes at wave vector ${\bf k}=(0,0)$. The corresponding Schrödinger equation can be solved by expanding the Hamiltonian in Eq.~\eqref{HCh_kspace} near ${\bf k}=(0,0)$ and then replacing $k_{i}\rightarrow-i\partial_{i}$. Considering an edge state whose energy dispersion is linear, i.e., $E(k_{x})=Ak_{x}$, we have 
\begin{eqnarray}
\left\{-iA\sigma_{x}\partial_{x}-iA\sigma_{y}\partial_{y}+\left[M-B\partial_{x}^{2}-B\partial_{y}^{2}\right]\sigma_z\right\}\psi
=Ak_{x}\psi.
\nonumber \\
\end{eqnarray}
Using the ansatz $\psi=\psi_{x}\psi_{y}\propto e^{ik_{x}x}e^{-\lambda y}\chi_{\eta}$, with $\chi_{\eta}$ a spinor of eigenvalue $\eta$, the equation becomes 
\begin{eqnarray}
\left\{Ak_{x}\sigma_{x}+iA\sigma_{y}\lambda+\left[M+Bk_{x}^{2}-B\lambda^{2}\right]\sigma_z\right\}\chi_{\eta}
=Ak_{x}\chi_{\eta}.
\nonumber \\
\label{eq:edge}
\end{eqnarray}
From the equation above, we see that when the edge state is an eigenstate of $\sigma_{x}\chi_{\eta}=\eta\chi_{\eta}=$ with eigenvalue $\eta=\pm 1$, the $Ak_{x}$ terms on the left and right hand sides cancel out. The remaining equation $A\lambda+\left(M+Bk_{x}^{2}-B\lambda^{2}\right)=0$, straightforwardly 
yields the edge state decay length $1/\lambda$. Note that the edge state is an eigenstate of $\sigma_{x}$ because we choose the boundary to be extending along the ${\hat{\bf x}}$ direction. We also remark that the topological state of the Chern insulator is protected by the approximate chiral symmetry of the corresponding nodal semimetal, which has a nonzero winding number, as discussed in detail in Ref.~[\onlinecite{Candido218}].

The continuum limit solution above being an eigenstate of $\sigma_x$ implies that in the lattice model the edge state at any site $i$ has equal weights of the two orbitals $\left\{s,p\right\}$ (pseudospin degrees of freedom). Note, however, that in this simple calculation we suppose a perfect linear dispersion $E(k_{x})=Ak_{x}$, which is not true for the edge states at large momenta $k_{x}\apprge M/A$. This means in reality that only the edge state at zero momentum $k_{x}=0$ is a perfect eigenstate of $\sigma_{x}$.

In what follows, we examine two different models for the NM layer to highlight the influence of the orbital kinetics of the NM on the fine-tuned zero-energy states, and whether they are distinguishable from the true edge states at zero momentum $k_{x}=0$. Here we consider a NM layer with parabolic bands. A Chern insulating layer coupled to a gapless metallic system, also described by Dirac models, has been considered previously~\cite{Baum15}.

\subsubsection{Results: Chern insulator/NM junction with identical orbital kinetics in the NM} 

We first consider an NM tight-binding model that has identical kinetic hopping terms for the $s$ and $p$ orbitals, i.e.,
\begin{eqnarray}
H_{NM}^{\sigma_{0}}&=&\sum_{i\in NM,\delta}t_{N}\left(c_{i,s}^{\dag}c_{i+\delta, s}+c_{i,p}^{\dag}c_{i+\delta, p}\right)+h.c.
\nonumber \\
&&-\sum_{i\in NM}\mu_{N}\,\left(c_{i,s}^{\dag}c_{i,s}+c_{i,p}^{\dag}c_{i,p}\right),
\label{HNM_sigma0}
\end{eqnarray}
with $t_N$ and $\mu_{N}$ the hopping and chemical potential, respectively. The coupling Hamiltonian reads
\begin{eqnarray}
H_{BD}^{\sigma_{0}}&=&\sum_{i\in BD}t_{B}\left(c_{i,s}^{\dag}c_{i+b,s}+c_{i,p}^{\dag}c_{i+b,p}\right)+h.c.,
\label{HBD_sigma0}
\end{eqnarray}
where $t_{B}$ is the interface hopping parameter and $i\in$BD denotes the boundary sites at the interface. We use the superscript $\sigma_{0}$ in $H^{\sigma_{0}}$ to denote that when the Fourier transform of such a Hamiltonian is written in the basis $(c_{{\bf k},s},c_{{\bf k},p})^{T}$, it is proportional to the identity matrix.

\begin{figure}[ht]
\begin{center}
\includegraphics[clip=true,width=0.99\columnwidth]{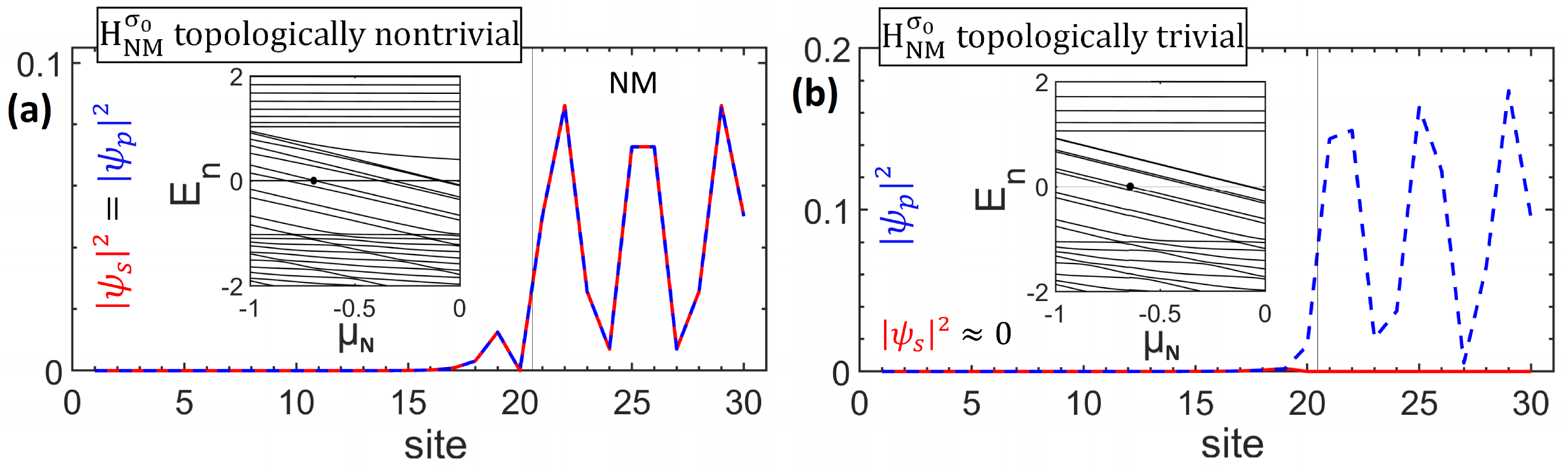}
\caption{Chern insulator/NM junction with periodic boundary conditions in ${\hat{\bf x}}$ and open boundary conditions in ${\hat{\bf y}}$. Here, the NM and interface Hamiltonians have {\it identical} kinetics between the two orbitals. Panel (a) shows the wave functions $|\psi_{s}(y)|^{2}$ (solid red curve) and $|\psi_{p}(y)|^{2}$ (dashed blue curve) for the zero-energy state in the topologically (a) nontrivial ($M=-1$) and (b) trivial ($M=1$) cases. The insets show the eigenenergies $E_n$ at $k_{x}=0$ as functions of $\mu_N$, and the black dots indicate the zero-energy states, whose wave functions are shown. We have used the parameters $t = t'= t_N = t_B = 1$.}
\label{fig:Chern_NM_junction}
\end{center}
\end{figure}

\textit{Nontrivial phase ($M<0$).} In the inset of Fig. \ref{fig:Chern_NM_junction} (a) we show the eigenenergies $E_n$ at $k_{x}=0$ versus $\mu_N$. Note that in addition to the topological edge state, pinned at zero energy, there are several trivial NM states within the Chern insulator bulk gap that eventually cross zero. This is similar to what happens in the SSH/NM system in Sec.~\ref{sec:SSHNM}.

As an example, we show in Fig.~\ref{fig:Chern_NM_junction} (a) the wave function of the fine-tuned zero-energy state at $k_{x}=0$ for $\mu_N=-0.6906t$ (black dot in the inset). We note that, similarly to the true topological edge state, the state here also has equal $s$- and $p$- orbital components ($|\psi_s|^2 = |\psi_p|^2$) in both the Chern insulator and NM regions. This indicates that it mimics the true topological edge state. 

\textit{Trivial phase ($M>0)$.} In Fig.~\ref{fig:Chern_NM_junction} (b) we plot $E_n$ versus $\mu_N$ (inset) and show the wave function of one of the fine-tuned zero-energy states $\mu_N=-0.6485t$ ( black dot in the energy spectrum.) In this case, the wave function no longer has the same weights on both orbitals ($|\psi_s|^2 \approx 0$, for this particular example), and hence in principle should be distinguishable from the true topological edge state.

\subsubsection{Results: Chern insulator/NM junction with opposite orbital kinetics in the NM}

The second NM model we analyze has kinetic hopping terms of opposite signs for the $s$ and $p$ orbitals, i.e., 
\begin{eqnarray}
H_{NM}^{\sigma_{z}}&=&\sum_{i\in NM,\delta}t_{N}\left(c_{i,s}^{\dag}c_{i+\delta, s}-c_{i,p}^{\dag}c_{i+\delta, p}\right)+h.c.
\nonumber \\
&&-\sum_{i\in NM}\mu_{N}\,\left(c_{i,s}^{\dag}c_{i,s}-c_{i,p}^{\dag}c_{i,p}\right).
\label{HNM_sigmaz}
\end{eqnarray}
For the interface, the Hamiltonian is given by
\begin{eqnarray}
H_{BD}^{\sigma_{z}}&=&\sum_{i\in BD}t_{B}\left(c_{i,s}^{\dag}c_{i+b, s}-c_{i,p}^{\dag}c_{i+b, p}\right)+h.c.
\label{HBD_sigmaz}
\end{eqnarray}
Here, the superscript $\sigma_{z}$ means that a Bloch Hamiltonian $H^{\sigma_z}(\mathbf{k})$, written in the basis $(c_{{\bf k},s},c_{{\bf k},p})^{T}$, is proportional to $\sigma_{z}$.

\begin{figure}[htb!]
\begin{center}
\includegraphics[clip=true,width=0.99\columnwidth]{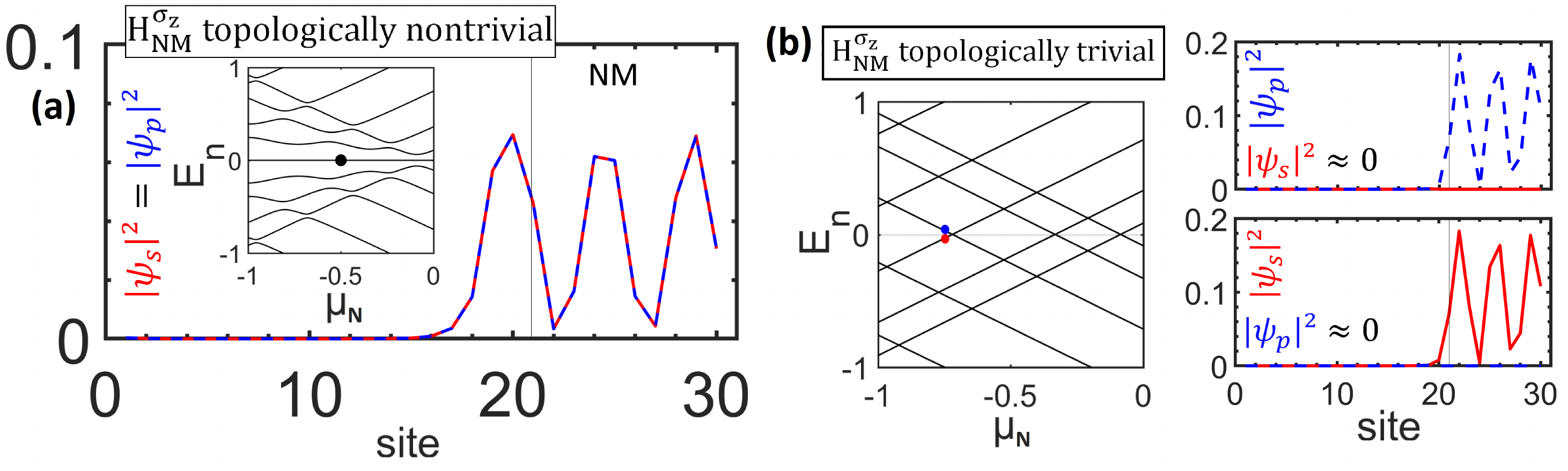}
\caption{Chern insulator/NM junction with periodic boundary conditions in ${\hat{\bf x}}$ and open boundary conditions in ${\hat{\bf y}}$. Here the NM and interface Hamiltonians have opposite kinetics between the two orbitals. The wave functions $|\psi_{s}(y)|^{2}$ (solid red curve) and $|\psi_{p}(y)|^{2}$ (dashed blue curve) for the zero-energy state are shown for the (a) nontrivial ($M=-1$) and (b) trivial ($M=1$) cases. In the trivial case the solutions come in pairs. The inset shows the eigenenergies $E_{n}$ as functions of $\mu_{N}$, with the parameters $t = t'= t_N = t_B = 1$.} 
\label{fig:Chern_NM_junction2}
\end{center}
\end{figure}

\textit{Nontrivial phase ($M<0$).} In Fig.~\ref{fig:Chern_NM_junction2}(a) we show that as we vary $\mu_N$, it is not possible to generate an additional zero-energy state. As a result, the true topological edge states can be unambiguously identified.
This is so because there are already two solutions (true topological edge states) at $E=0$. 
%Notice that, even if one of the zero-modes was shifted away from zero, crossings would be prohibited since the system is PH symmetric (cf. trivial case in Fig.~\ref{fig:Chern_NM_junction2} (b)). 
%As we have mentioned previously~\ref{footnote1}, our 1D system allows for at most two solutions per energy. 

\textit{Trivial phase ($M>0$).} In this case, as shown in Fig.~\ref{fig:Chern_NM_junction2}(b), it is always possible to create zero-energy states that come in pairs. In a way similar to the Majorana chain/NM juncton discussed in Sec.~\ref{Kitaev:trivial}, one of the two zero-energy states is purely an $s$-like wave function and the other is a purely $p$-like, making two of them together indistinguishable from the true topological edge state that has equal $s$ and $p$ orbital wave functions.

\section{Conclusions}

In summary, we show that when an NM is attached to a topological material, it is possible to fine tune the chemical potential of the metallic layer so as to create a zero-energy state that is indistinguishable from the true topological edge state. The results for the three examples we have examined are summarized below.

For the SSH/NM junction, we find that such a zero-energy state can be created by fine-tuning the chiral-symmetry breaking chemical potential of the NM. Moreover, these states can be created in both the topologically trivial and nontrivial phases, and exist over a wide region of the topological phase diagram. The wave function profile of this zero-energy state in the SSH region is only localized on one sublattice, just like the true topological edge state, making it indistinguishable from the true topological edge state as far as the wave function profile is concerned.

For the Kitaev/NM junction in the topologically nontrivial phase, we find that such a zero-energy state cannot occur alongside the true Majorana edge states. On the other hand, in the trivial phase, the fine-tuded zero energy states can occur, and they must appear in pairs due to PH symmetry. Because one of them has an electron-like wave function and the other a hole-like, it is highly possible to mistaken the two zero-energy states together as one single Majorana fermion. In addition, though investigating the local and global DOS, we find that it is practically impossible to distinguish these fine-tuned states from true Majorana fermions either by detecting the local DOS using STM, or by probing the tunneling conductance of the whole junction, since these fine-tuned states give the same zero-bias peak as the true Majorana fermions.

In the 2D Chern insulator, the true topological edge state self-generates a symmetry eigenvalue due to the direction of the edge. As a result, its wave function has the same weight on the two orbitals. In the 2D Chern insulator/NM junction, we show that for an NM model that has the same kinetic hopping terms for the two orbital degrees of freedom, a zero-energy state can be created in the topologically nontrivial phase. The wave function, here, has the same weights on the two orbitals, thus mimicking the true topological edge state. On the other hand, if the kinetic hopping terms of the two orbitals in the NM have opposite signs, then the zero-energy state appear in pairs and can only be created in the topologically trivial phase. Moreover, one of the zero-energy states has an $s$-like wave function and the other a $p$-like, meaning that the two states together may be wrongly identified as one single edge state. 

These results indicate that should a metallic layer be attached to a topological material, caution must be taken if one intends to identify the true topological edge state merely from the zero-energy modes and their wave function profiles, or some zero-bias conductance feature in STM or tunneling conductance measurements, since such fine-tuned zero-energy states may occur in a wide region of both the topologically trivial and nontrivial phases.

\begin{acknowledgments}

I.C. and P.H.P contributed equally to this work. P.H.P acknowledges support of the PNPD program by Coordena\c c\~ao de Aperfei\c coamento de Pessoal de N\'{i}vel Superior (CAPES) –- Finance Code 001. J.C.E. acknowledges support from the S\~ao Paulo Research Foundation (FAPESP) Grants No. 2016/08468-0 and
No. 2020/00841-9, and from Conselho Nacional de Pesquisas (CNPq), Grant No. 306122/2018-9. 

\end{acknowledgments}

\bibliography{Literatur}

\end{document}